\documentclass[prl,%superscriptaddress,
twocolumn,letterpaper,%
nofootinbib,showpacs,preprintnumbers,amsmath,amssymb]{revtex4}
\usepackage{bm,curves}

\newcommand{\footfrac}[2]%

\def\cxx{1 - {{4x^{2} P^{2}}\over{Q^{2}}} }
\def\cx{1 - {{2x    P^{2}}\over{Q^{2}}} }

\def\cut{\sqrt{1 - {{4(m^{2}+\lambda^2)}\over {s}} } }

\begin{document}

\title{
Transverse Momentum in Semi-Inclusive Polarized Deep Inelastic 
Scattering \\
and the Spin-Flavor Structure of the Proton
}
\author{Steven D. Bass}
\affiliation{
High Energy Physics Group, 
Institute for Experimental Physics and
Institute for Theoretical Physics, Universit\"at Innsbruck,
Technikerstrasse 25, A 6020 Innsbruck, Austria}

\begin{abstract}
\noindent
The non-valence spin-flavor structure of the proton extracted from 
semi-inclusive measurements of polarized deep inelastic scattering 
depends strongly on the transverse momentum of the detected hadrons 
which are used to determine the individual polarized sea distributions.
This physics may explain the recent HERMES observation of a
positively polarized strange sea through semi-inclusive scattering, 
in contrast to the negative strange sea polarization deduced from 
inclusive polarized deep inelastic scattering.
\end{abstract}
\maketitle

\vfill\eject
%
%\section{Introduction}
%

Understanding the internal spin structure of the proton is one of the 
most challenging problems facing subatomic physics: 
How is the spin of the proton built up out from the intrinsic spin 
and orbital angular momentum of its quark and gluonic constituents ? 
A key issue is the contribution of polarized sea quarks 
(up, down and strange) in building up the spin of the proton.
Fully inclusive polarized deep inelastic scattering experiments 
suggest that the sea carries significant {\it negative} polarization
(polarized in the opposite direction to the spin of the proton) \cite{windm}. 
In contrast, new semi-inclusive measurements performed by the HERMES 
\cite{hermessemi,miller} and SMC \cite{smcsemi} 
experiments with final state particle identification 
suggest that the light-flavored (up and down) 
sea contributes 
close to zero to the proton's spin, and that 
the strange sea is {\it positively} polarized.

Here we argue that the transverse momentum of the detected final-state 
hadrons in semi-inclusive scattering may be essential to understanding 
these results.
The first moment of the $g_1$ spin structure function for polarized 
photon-gluon fusion $(\gamma^* g \rightarrow q {\bar q})$
receives a positive contribution proportional to the mass squared of 
the struck quark or antiquark which originates from low values of 
quark transverse momentum, $k_t$, 
with respect to the photon-gluon direction.
It also receives a negative contribution 
from $k_t^2 \sim Q^2$, where $Q^2$ is the virtuality of the hard 
photon
--- see Eqs~(\ref{eq:7}-\ref{eq:13}) below.
Thus, the spin-flavor structure of the sea extracted from semi-inclusive 
measurements depends strongly on the $k_t$ distribution of the detected 
hadrons.
Understanding this physics is essential to ensure that 
the theory and experimental acceptance are correctly matched 
when extracting new information from present and future experiments.
We first give a brief overview of the experimental situation and 
then explain the importance of transverse momentum, 
including calculations in the kinematics of the HERMES and SMC experiments.

Following pioneering experiments at SLAC \cite{slac}, recent experiments 
in fully inclusive polarized deep inelastic scattering have extended 
measurements of the nucleon's $g_1$ spin dependent structure function 
to lower values of Bjorken $x$ where the nucleon's sea becomes important
\cite{windm}.
From the first moment of $g_1$,
these experiments have revealed 
a small value for the flavour-singlet axial-charge:
\begin{equation}
g_A^{(0)}\bigr|_\mathrm{pDIS} 
= \Delta u + \Delta d + \Delta s 
= 0.2 - 0.35 .
\label{inv}\end{equation}
This result is particularly interesting \cite{spinrev}
because $g_A^{(0)}$ 
is interpreted in the parton model as the fraction of the proton's 
spin which is carried by the intrinsic spin of its quark and antiquark 
constituents.
The value (\ref{inv}) is about half the prediction of relativistic 
constituent quark models ($\sim 60\%$).
It corresponds to a negative strange-quark polarization
\begin{equation}
\Delta s = -0.10 \pm 0.04 .
\label{deltas}
\end{equation}

The small value of $g_A^{(0)}$ measured in polarized deep inelastic
scattering 
has inspired vast experimental and theoretical activity to understand the 
spin structure of the proton.
New experiments are underway or being planned to map out 
the proton's spin-flavor structure and to measure 
the amount of spin carried by polarized gluons in the polarized proton. 
These include semi-inclusive polarized deep inelastic scattering
\cite{miller}, 
polarized proton-proton collisions at the Relativistic Heavy Ion 
Collider (RHIC) \cite{rhic}, and polarized $ep$ collider studies 
\cite{bassdr}.

Here we focus on semi-inclusive measurements of fast pions and kaons 
in the current fragmentation region. 
These measurements are being used by the HERMES experiment 
\cite{hermessemi,miller}
(following earlier work by SMC \cite{smcsemi}) 
to reconstruct the individual up, down and strange 
quark contributions to the proton's spin.
In contrast to inclusive polarized deep inelastic scattering 
where the $g_1$ structure function is deduced by detecting only 
the scattered lepton \cite{close}, 
the detected particles in the semi-inclusive experiments are 
high-energy (greater than 20\% of the energy of the incident photon)
charged pions and kaons in coincidence with the scattered lepton.
For large energy fraction $z=E_h/E_{\gamma} \rightarrow 1$
the most probable occurence is that
the detected $\pi^{\pm}$ and $K^{\pm}$
contain the struck quark or antiquark in their valence Fock state. 
They therefore act as a tag of the flavor of the struck quark \cite{close}.
In leading order (LO) QCD:
\begin{equation}
A_{1p}^{h} (x,Q^2) \simeq
{
\sum_{q,h} e_q^2 %\biggl[ 
\Delta q(x,Q^2) \int_{z_{\rm min}}^1 D_q^h (z, Q^2) 
\over
\sum_{q,h} e_q^2 %\biggl[ 
q(x,Q^2) \int_{z_{\rm min}}^1 D_q^h (z, Q^2)
}
\label{semia1}
\end{equation}
where $z_{\rm min} \sim 0.2$.
Here
\begin{equation}
D_q^h (z, Q^2) = \int d k_t^2 D_q^h (z, k_t^2, Q^2)
\end{equation}
is the fragmentation function for the struck quark or antiquark 
to produce a hadron $h$ ($=\pi^{\pm}, K^{\pm}$) carrying energy 
fraction $z=E_h/E_{\gamma}$ in the target rest frame;
$\Delta q(x,Q^2)$ 
is the quark (or antiquark) 
polarized parton 
distribution and $e_q$ is the quark charge.
Note the integration over the transverse momentum $k_t$ of the 
final-state hadrons \cite{closem}. 
(In practice this integration over $k_t$ is determined 
 by the acceptance of the experiment.)
Since pions and kaons have spin zero, the fragmentation functions 
are the same for both polarized and unpolarized leptoproduction.
NLO corrections to Eq.~(\ref{semia1}) are discussed in Ref.~\cite{defl}.

The semi-inclusive data reported by HERMES and SMC suggest that the 
(measured) light-flavored (up and down quark) sea contributes close 
to zero to the spin of the proton \cite{hermessemi,miller,smcsemi}
and that 
the (measured) strange sea polarization is {\it positive}
\cite{miller},
in contrast 
with the result (\ref{deltas}) deduced from inclusive scattering.
The mean $Q^2$ for these experiments is 2.5GeV$^2$ (HERMES) and 
10GeV$^2$ (SMC).
For HERMES the average transverse momentum of 
the detected fast hadrons is less than about 0.5 GeV \cite{hermespriv},
whereas for SMC the $k_t$ of the detected 
fast pions was less than about 1 GeV \cite{smcpriv}.
Further semi-inclusive measurements are planned with 
the COMPASS experiment at CERN and 
the proposed future Electron-Ion Collider (EIC), 
extending the kinematic range to smaller Bjorken $x$.

Transverse momentum is essential to understanding polarized semi-inclusive 
data.
Consider the polarized photon-gluon fusion process 
$\gamma^* g \rightarrow q {\bar q}$.
We evaluate the $g_1$ spin structure function for this process as a 
function of the transverse momentum squared of the struck quark, $k_t^2$, 
with respect to the photon-gluon direction.
We use $q$ and $p$ to denote the photon and gluon momenta and
use the cut-off $k_t^2 \geq \lambda^2$ to separate the total 
phase space into ``hard'' ($k_t^2 \geq \lambda^2$) and ``soft'' 
($k_t^2 < \lambda^2$) contributions.
One finds \cite{bnt,bassbs}:
\begin{widetext}
\begin{eqnarray}
& & 
g_1^{(\gamma^* g)}|_{\rm hard} = 
%(x,Q^2) =
\nonumber \\
& &
-{\alpha_s  \over 2 \pi }
{\cut \over \cxx} \Biggl[ (2x-1)(\cx) 
\Biggl\{1 
- {1 \over {\cut \sqrt{\cxx} }}
\ln \biggl({ {1+\sqrt{\cxx} \cut}\over {1-\sqrt{\cxx} \cut}}
\biggr) 
\Biggr\} 
\nonumber \\
& &
\ \ \ \ \ \ \ \ \ \ \ \ \ \ \ \ \ \ \ \ \ \ \ \ \ \ \ \ \ \
+ (x-1+{{x P^{2}}\over{Q^{2}}})
{{\left( 2m^{2}(\cxx)- P^{2}x(2x-1)(\cx)\right)}
\over {(m^{2} + \lambda^2) (\cxx) - P^{2}x(x-1+{{x P^{2}}\over{Q^{2}}})}}
\Biggr]
\label{eq:7}
\end{eqnarray}
for each flavor of quark liberated into the final state.
Here 
$m$ is the quark mass, $Q^2 =-q^2$ is the virtuality of the hard photon, 
$P^2=-p^2$ is the virtuality of the gluon target,
$x$ is the Bjorken variable ($x= {Q^2 \over 2 p.q}$) 
and 
$s$ is the centre of mass energy squared,
$
s= (p+q)^2 = Q^2 \bigl( {1 - x \over x} \bigr) - P^2
$,
for the photon-gluon collision.
When $Q^2 \rightarrow \infty$ 
the expression for $g_1^{(\gamma^* g)}|_{\rm hard}$ 
simplifies to the leading twist (=2) contribution:
\begin{equation} 
g_1^{(\gamma^* g)}|_{\rm hard}
= {\alpha_s \over 2 \pi} \Biggl[ (2x-1) \Biggl\{
\ln {Q^2 \over \lambda^2} + \ln {1-x \over x} - 1 
+ \ln {\lambda^2 \over {x(1-x) P^2 + (m^2 + \lambda^2)} } \Biggr\}
+
(1 -x) { {2m^2 - P^2x(2x-1)} \over { m^2 + \lambda^2 - P^2 x(x-1)} }
\Biggr] .
\label{eq:11}
\end{equation}
Here we take $\lambda$ to be independent of $x$.
Note that for finite quark masses, 
phase space limits Bjorken $x$ to
$x_{max} = Q^2 / (Q^2 + P^2 + 4 (m^2 + \lambda^2))$
and protects 
$g_1^{(\gamma^* g)}|_{\rm hard}$ 
from reaching the $\ln (1-x)$ singularity in Eq. (\ref{eq:11}).  
For this photon-gluon fusion process, 
the first moment of the ``hard'' contribution is \cite{bnt}:
\begin{equation}
\int_0^1 dx g_1^{(\gamma^{*} g)}|_{\rm hard} 
= - {\alpha_s \over 2 \pi} 
\left[1 + \frac{2m^{2}}{P^{2}}
\frac{1}{\sqrt{1+4(m^{2}+\lambda^2)/P^{2}}} 
\ln \left(
\frac{\sqrt{1+4(m^{2}+\lambda^2)/P^{2}} -1}
{\sqrt{1+4(m^{2}+\lambda^2)/P^{2}} +1} \right) \right].
\label{eq:12}
\end{equation}
The ``soft'' contribution to the first moment of $g_1$ is then
obtained by subtracting Eq. (\ref{eq:12})
from the inclusive first moment (obtained by setting $\lambda =0$):
\begin{eqnarray}
& & \int_0^1 dx g_1^{(\gamma^{*} g)}|_{\rm soft} = 
\nonumber \\
& & {\alpha_s \over 2 \pi} \frac{2m^2}{P^2}
\left[
\frac{1}{\sqrt{1+4(m^2+\lambda^2)/P^2}} 
\ln \left(
\frac{\sqrt{1+4(m^2 + \lambda^2)/P^2} -1}
{\sqrt{1+4(m^2 +\lambda^2)/P^2} +1} \right) 
-
\frac{1}{\sqrt{1+4 m^2 / P^2}} 
\ln \left(
\frac{\sqrt{1+4 m^2 /P^2} -1}{\sqrt{1+4 m^2 /P^2} +1} \right) 
\right].
\label{eq:13}
\end{eqnarray}
\end{widetext}
Eq.~(\ref{eq:13}) measures the contribution to the polarized sea from 
$k_t$ less than the cut-off $\lambda$ in the limit $Q^2 \rightarrow \infty$.

For fixed gluon virtuality $P^2$ the photon-gluon fusion process
induces two distinct contributions to the first moment of $g_1$.
Consider the leading twist contribution, Eq. (\ref{eq:12}).
The first term, $-{\alpha_s \over 2 \pi}$, in Eq.(\ref{eq:12}) 
is mass-independent and comes from the region of phase space 
where the struck quark carries large transverse momentum squared 
$k_t^2 \sim Q^2$.
It measures a contact photon-gluon interaction and is associated 
\cite{ccm,bint}
with the 
axial anomaly \cite{adler}.  
The second mass-dependent term comes from the region of phase-space 
where the struck quark carries transverse momentum $k_t^2 \sim m^2,P^2$.  
This {\it positive} mass dependent term is proportional to the mass 
squared of the struck quark.
The mass-dependent in Eqs.~(\ref{eq:12}) and (\ref{eq:13}) 
can safely be neglected for light-quark flavor (up and down) production.
It is very important for strangeness (and charm \cite{bbs2,fms}) 
production.
For vanishing cut-off ($\lambda^2=0$) this term vanishes in the limit 
$m^2 \ll P^2$ and tends to $+{\alpha_s \over 2 \pi}$ when $m^2 \gg P^2$
(so that the first moment of $g_1^{(\gamma^* g)}$
 vanishes in this limit)
\footnote{
Note that there is no contribution to the first moment 
from the intermediate region of $k_t^2$, 
between the soft scale set by $m^2$ and $P^2$ and the 
hard scale set by $Q^2$, because the first moment of 
the spin-dependent Altarelli-Parisi splitting function 
for $g \rightarrow q {\bar q}$ vanishes: $\int_0^1 dx (2x-1) = 0$.
}
.

Eq. (\ref{eq:12}) leads to the well known formula \cite{etar,ccm,bint}
\begin{equation}
g_A^{(0)}|_{\rm pDIS} = 
\Biggl( \sum_q \Delta q 
  - 3 {\alpha_s \over 2 \pi} \Delta g \Biggr)_{\rm partons} %%+ \ {\cal C}
\label{eq:18}
\end{equation}
where $\Delta g$ is the amount of spin carried by polarized gluon 
partons in the polarized proton and
$\Delta q_{\rm partons}$ measures 
the spin carried by quarks and antiquarks
carrying ``soft'' transverse momentum $k_t^2 \sim m^2, P^2$, 
with
$P^2$ a typical gluon virtuality in the proton wavefunction.
Since $\Delta g \sim 1/{\alpha_s}$ under QCD evolution \cite{etar}, 
the two terms in Eq. (\ref{eq:18}) both scale as $Q^2 \rightarrow \infty$.

\begin{figure}[h] 
\includegraphics{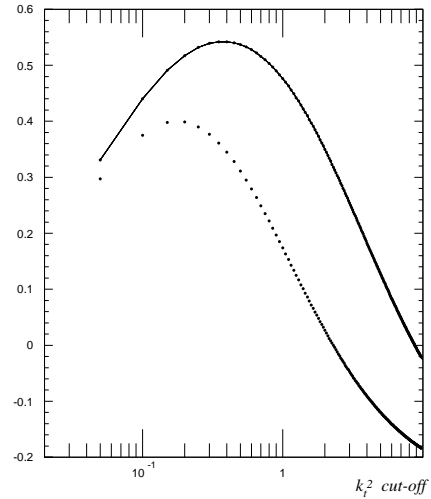} %\label{f1a}
\begin{center} 
\vspace{6.5cm} 
\parbox{8.0cm} 
{\caption[Delta]
{
$\int_0^1 dx \ g_1^{(\gamma^* g)}|_{\rm soft}$ 
for polarized strangeness production 
with $k_t^2 < \lambda^2$
in units of ${\alpha_s \over 2 \pi}$.
Here 
$Q^2=2.5$GeV$^2$ (dotted line) and 10GeV$^2$ (solid line).
}
\label{fig1}} 
\end{center} 
\end{figure}
\begin{figure}[h] 
\includegraphics{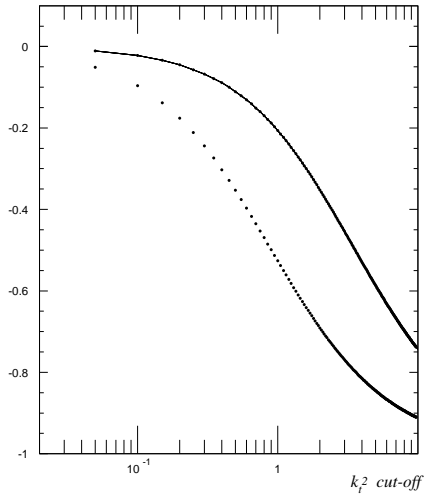} %\label{f1a}
\begin{center} 
\vspace{6.5cm} 
\parbox{8.0cm} 
{\caption[Delta]
{
$\int_0^1 dx \ g_1^{(\gamma^* g)}|_{\rm soft}$ 
for light-flavor ($u$ or $d$) production 
with $k_t^2 < \lambda^2$ 
in units of ${\alpha_s \over 2 \pi}$.
Here 
$Q^2=2.5$GeV$^2$ (dotted line) and 10GeV$^2$ (solid line).
}
\label{fig2}} 
\end{center} 
\end{figure}
We now evaluate the effect of the strange quark mass on the $k_t$ 
distribution of the final-state hadrons produced in photon-gluon
fusion.
In Figs. 1 and 2 we show the first moment of 
$g_1^{(\gamma^* g)}|_{\rm soft}$ for the strange and light 
(up and down) flavors respectively as a function of the 
transverse momentum cut-off (acceptance) $k_t^2 < \lambda^2$.
Here we set $Q^2 =2.5$GeV$^2$ 
(corresponding to the HERMES experiment) 
and 10GeV$^2$ (SMC)
and integrate the full expression in Eq.~(\ref{eq:7}).
Following \cite{ccm}, 
we take $P^2 \sim \Lambda_{\rm qcd}^2$ and set $P^2 = 0.1$GeV$^2$.
We take the strange quark mass $m=0.2$GeV.
Observe the small value for the light-quark 
polarization at low transverse momentum and 
the positive value for the integrated 
strange sea polarization at low $k_t^2$:
$k_t < 1.5$GeV at the HERMES $Q^2=2.5$GeV$^2$.
When we relax the cut-off, increasing the acceptance of the experiment,
the measured strange sea polarization changes sign and becomes 
negative (the result implied by fully inclusive deep inelastic measurements).

For $P^2=0.1$GeV$^2$ the fully inclusive first moment of 
$g_1^{(\gamma^* g)}$ 
for strange quark production is 
0.26 times $-{\alpha_s \over 2 \pi}$ at $Q^2=2.5$GeV$^2$,
and 
0.28 times $-{\alpha_s \over 2 \pi}$ at $Q^2=10$GeV$^2$ 
(and also in the scaling limit $Q^2\rightarrow \infty$).
In practice a full calculation of the $k_t$ dependence of 
the polarized sea would involve integrating over 
the distribution of gluon virtualities 
in the proton wavefunction \cite{bbs2}.
While this distribution is strongly peaked at small $P^2$, 
it is interesting to investigate the effect of increasing 
the value of $P^2$.
For $Q^2=2.5$ (10) GeV$^2$
the contribution to the total polarized strangeness 
from $k_t < \lambda$ changes sign and becomes 
negative at $\lambda = 0.7$ (1.4) GeV for a gluon 
with $P^2=0.5$GeV$^2$ 
and 
$\lambda = 0.5$ (1.0) GeV
for a highly virtual gluon with $P^2=1.0$GeV$^2$.
For $Q^2=2.5$ (10) GeV$^2$
the total, inclusive, 
polarized strangeness
$\int_0^1 dx \ g_1^{(\gamma^* g)}$
induced 
by photon-gluon fusion
increases 
to 0.52 (0.61)
times $-{\alpha_s \over 2 \pi}$ for $P^2=0.5$GeV$^2$
and
0.54 (0.71)
times $-{\alpha_s \over 2 \pi}$ for $P^2=1.0$GeV$^2$.

We summarize our results.
The small value for the light-quark sea polarization and the positive 
strange sea polarization observed in semi-inclusive measurements of 
polarized deep inelastic scattering may have a simple interpretation 
in terms of the transverse momentum dependence of polarized photon-gluon 
fusion.
It would be very interesting to extend the present measurements 
to include final-state hadrons with the highest values of transverse 
momentum, $k_t$,
to look for the growth in the (negative) polarization of 
the light-(up and down) flavor sea and the sign change 
for the polarized strangeness that are expected at large $k_t$.
Finally, we note that an independent 
measurement of 
the strange quark axial-charge, $\Delta s$, 
could be obtained from a precision measurement of elastic $\nu p$ scattering
\cite{nup}.
(The elastic $\nu p$ process is independent of 
 assumptions about the behaviour of spin structure functions at $x \sim 0$).
Semi-inclusive and inclusive polarized deep inelastic scattering 
together with
elastic $\nu p$ scattering 
provide complementary information about the transverse momentum and 
Bjorken $x$ distributions of strange quark polarization in the nucleon.

\vspace{0.5cm}

\acknowledgments
SDB is supported by a Lise Meitner Fellowship, M683, from the Austrian FWF. 
It is a pleasure to thank A.W. Thomas for stimulating discussions and 
B. Badelek, A. Deshpande, N. Makins and A. Miller for helpful communications 
about experimental data.

%\newpage

\newpage

\end{document}